\documentclass[12pt]{article}
\usepackage{tabularx}
\usepackage{array}
\usepackage{graphics}
\usepackage{graphicx}
\usepackage{epsfig}
\usepackage{epsf}
\usepackage{amssymb}
\usepackage{citesort}
\usepackage{color}
\makeatletter

\usepackage{verbatim}

\newcommand{\msbar}{{\overline{\rm MS}}}
\newcommand{\ri}{{\rm RI/MOM}}

\newcommand{\bea}{\begin{eqnarray}}
\newcommand{\eea}{\end{eqnarray}}
\newcommand{\beq}{\begin{equation}}
\newcommand{\eeq}{\end{equation}}

\newcommand{\ga}{{\gamma}}
\newcommand{\1}{{1\hspace*{-.272cm}1}}
\newcommand{\Z}{{Z\hspace*{-.232cm}Z}}
\newcommand{\nn}{\nonumber}

\newcommand{\pdir}{p\kern -5.2pt\raise 0.2ex\hbox {/}}
\newcommand{\vdir}{v\kern -5.75pt\raise 0.15ex\hbox {/}}
\newcommand{\kdir}{k\kern -5.75pt\raise 0.15ex\hbox {/}}
\newcommand{\epsdir}{\epsilon\kern -5.0pt\raise 0.15ex\hbox {/}}
\newcommand{\bvdir}{\bar{v}\kern -5.75pt\raise 0.15ex\hbox {/}}
\newcommand{\Ddir}{D\kern -7.75pt\raise 0.20ex\hbox {/}}
\newcommand{\ldir}{l\kern -5.0pt\raise 0.2ex\hbox{/}}
\newcommand{\varepsdir}{\varepsilon\kern -5.5pt\raise 0.15ex\hbox{/}}

\newcommand{\kkbar}{K^0\!-\!{\overline{K^0}}}

\def\l{\left}
\def\r{\right}
\def\nn{\nonumber}

\makeatother

\begin{document}

\thispagestyle{empty}
\begin{flushright}
\begin{tabular}{l}
{\tt  CERN-PH-TH/2004-135}\\
{\tt  LPT Orsay 04-57}\\
{\tt  Roma-1384/04}\\                                              
\end{tabular}
\end{flushright}
\vskip 3.6cm\par
\begin{center}
{\par\centering \textbf{\Large Remarks on the hadronic matrix elements relevant } }\\
\vskip .25cm\par
{\par\centering \textbf{\Large to the SUSY $\kkbar$ mixing amplitude} }\\
\vskip 2.35cm\par
\scalebox{.9}{\par\centering \large
\sc Damir~Be\'cirevi\'c$^a$ and Giovanni~Villadoro$^{b}$}
\end{center}
{\par\centering \textsl{$^a$Laboratoire de Physique Th\'eorique (B\^at.210), Universit\'e Paris Sud}\\
\textsl{Centre d'Orsay, F-91405 Orsay-Cedex, France.}\\
\vskip 0.3cm\par} 
{\par\centering \textsl{$^b$Dip.~di Fisica, Univ.~di Roma ``La
Sapienza'',
}\\
\textsl{INFN, Sezione di Roma, P.le A.~Moro 2, I-00185 Roma,
Italy and}\\
\textsl{ CERN, Theory Division, CH-1211, Geneva 23, Switzerland.}\\
\vskip 2.7cm\par } 
\begin{abstract}
{We compute the $1$-loop chiral corrections to the bag parameters which 
are needed for the discussion of the SUSY $\kkbar$ mixing problem in 
both finite and infinite volume. We then show how the bag parameters can be combined 
among themselves and with some auxiliary quantities and thus sensibly reduce the 
systematic errors due to chiral extrapolations as well as those due to finite 
volume artefacts that are present in the results obtained from lattice QCD.  
We also show that in some cases these advantages remain as such even after 
including the $2$-loop chiral corrections. 
Similar discussion is also made for the $K\to \pi$  electro-weak penguin 
operators.}
\end{abstract}
\vskip 1.4cm

\setcounter{page}{1}
\setcounter{footnote}{0}
\setcounter{equation}{0}                                                        

\noindent
 
\renewcommand{\thefootnote}{\arabic{footnote}}
 
\newpage
\setcounter{page}{1}
\setcounter{footnote}{0}
\setcounter{equation}{0}
\section{Introduction}

We are entering the era of the large scale unquenched numerical simulations of QCD on the lattice 
and so the error on the $B_K$ parameter, currently dominated by the systematic error due to quenching~\cite{latt03}, 
is likely to fall below $10$\% quite soon. This will further improve our knowledge on 
the shape of the CKM unitarity triangle~\cite{UTA}, i.e., on the value of the CKM phase which 
is responsible for all the CP-violating phenomena in the Standard Model (SM). Since the CP-violation 
in SM is too small to explain the dynamical generation of the baryon asymmetry of the Universe~\cite{sakharov}, 
one is tempted to look for additional sources of CP-violation beyond the SM. Supersymmetric (SUSY) 
extensions of the SM, besides providing an elegant solution to the hierarchy problem, also provide 
new CP-violating phases whose size can be constrained by the experimentally measured processes 
governed by the flavor changing neutral currents (FCNC). 
A convenient way to study those is by using the mass insertion approximation~\cite{mia}.
In the basis in which the couplings of quarks and squarks to the neutral gauginos are flavor diagonal, 
the flavor changing insertions arise from the small off-diagonal terms in the squark masses, 
parameterized by dimensionless complex parameters
\bea
(\delta^{u/d}_{ij})_{AB} ={(\Delta_{ij}^{u/d})_{AB}\over m_{\tilde q}^2}\,,
\eea
where $m_{\tilde q}$ is the diagonal squark mass (averaged as $m_{\tilde q} = \sqrt{m_{\tilde q_i} m_{\tilde q_j}}$), 
and $(\Delta_{ij}^{u/d})_{AB}$ are the off-diagonal elements which mix both the left-handed and right-handed squark 
flavors ($i,j=1,2,3$; $A,B=L,R$). Like in the SM,  $\varepsilon_K$, the  parameter which 
measures the indirect CP-violation in the neutral kaon system, is given by
\bea
\varepsilon_K ={{\rm Im}\langle K^0 \vert {\cal H}^{\Delta S=2}_{\rm eff}\vert \bar K^0\rangle \over \Delta m_K\;\sqrt{2}}\,,
\eea
but unlike in the SM, the effective Hamiltonian in SUSY [${\cal H}^{\Delta S=2}_{\rm eff} = \sum_i C_i(\mu)O_i(\mu)$], 
besides the left-left ($LL$) four-quark operator, also contains the $LR$, $RL$ and $RR$ ones (see below for the specific bases of 
such operators). The Wilson coefficients, $C_i(\mu)$, can be computed perturbatively in any specific low energy SUSY model. 
The matrix elements, instead, must be computed non-perturbatively. To date the most suitable tool to do such a computation 
is by means of lattice QCD. The matrix elements of the non-SM operators are enhanced with respect to the SM one 
by a large factor, $m_K^2/(m_s+m_d)^2 \approx 25$, and therefore their more accurate determination is mandatory. 
So far these matrix elements have been computed in quenched lattice QCD by considering kaons consisting of degenerate 
quarks~\cite{allton}. The corresponding results are then used to discuss the constraints on $(\delta^d_{12})_{AB}$~\cite{susy-kkbar}, 
along the lines proposed in ref.~\cite{gabbiani}. In addition to unquenching, the lattice estimates should 
be improved by fixing one of the kaons' valence quarks to the physical strange quark mass (accessible in the lattice simulations) and 
pushing the other one as close to the chiral limit as possible. Due to limited computing resources, however, two problems appear: 
(1) one cannot work with the light quark as light as the physical $d$-quark and therefore a chiral extrapolation will always 
be needed; (2) the finite volume effects become more pronounced as the light quark mass is decreased. In such a situation 
and in order to get the 
phenomenologically interesting results  from lattice QCD on the matrix elements relevant to the SUSY $\kkbar$ mixing problem, one 
 should find a way to reduce uncertainties related to these two problems. In this paper these 
issues are addressed and further considered by using chiral perturbation theory (ChPT). We compute the chiral logarithmic corrections 
to the so called ``bag"-parameters and discuss the possible strategies that would allow one to minimize their impact onto 
the chiral extrapolations of the lattice results, as well as to minimize the systematic errors due to finite volume.
In Sec.~\ref{sec:1} we recall the frequently employed bases of $\Delta S=2$ operators and define the 
bag-parameters, to which we compute the chiral corrections in Sec.~\ref{sec:2}. In Sec.~\ref{sec:3} we discuss 
the combinations of bag parameters which are (completely or partially) free of chiral logarithms, and we briefly comment 
(Sec.~\ref{sec:4}) on the application of a similar strategy to the $K\to \pi$ matrix elements that are relevant 
to the $\Delta I=3/2$ amplitude in $K\to\pi\pi$ decay. In Sec.~\ref{sec:6} we discuss the impact of the $2$-loop 
chiral corrections to one class of the combinations of bag parameters, and we briefly conclude in Sec.~\ref{sec:5}.

\section{Bases of $\Delta S=2$ operators and $B$-parameters\label{sec:1}}

The SUSY contributions to the $\kkbar$ mixing amplitude are usually discussed 
in the so called SUSY basis of $\Delta S=2$ operators~\cite{gabbiani}:
\bea
 \label{baseS}
{\phantom{{l}}}\raisebox{-.16cm}{\phantom{{j}}}
O_1 &=& \ \bar s^i \gamma_\mu (1- \gamma_{5} )  d^i \,
 \bar s^j  \gamma_\mu (1- \gamma_{5} ) d^j \,  , 
  \nonumber \\
{\phantom{{l}}}\raisebox{-.16cm}{\phantom{{j}}}
O_2&=& \ \bar s^i  (1- \gamma_{5} ) d^i \,
\bar s^j  (1 - \gamma_{5} )  d^j \, ,  \nonumber  \\
{\phantom{{l}}}\raisebox{-.16cm}{\phantom{{j}}}
O_3&=& \ \bar s^i  (1- \gamma_{5} ) d^j \,
 \bar s^j (1 -  \gamma_{5} ) d^i \, ,  \\
{\phantom{{l}}}\raisebox{-.16cm}{\phantom{{j}}}
O_4 &=& \ \bar s^i  (1- \gamma_{5} )  d^i \,
 \bar s^j   (1+ \gamma_{5} ) d^j \,  ,  \nonumber \\
{\phantom{{l}}}\raisebox{-.16cm}{\phantom{{j}}}
O_5 &=& \ \bar s^i  (1- \gamma_{5} )  d^j \,
 \bar s^j   (1+ \gamma_{5} ) d^i \,  ,  \nonumber 
 \eea
where the superscripts stand for the color indices. Other bases are also employed, 
of which the Dirac basis is probably the most popular among the lattice QCD practitioners, 
\bea \label{baseF}
{\phantom{{l}}}\raisebox{-.16cm}{\phantom{{j}}}
&&Q_1 = 
\bar s^i \ga_\mu (1 - \ga_5) d^i\ \bar s^j  \ga_\mu (1 - \ga_5)d^j 
\;, \nonumber \\
{\phantom{{l}}}\raisebox{-.16cm}{\phantom{{j}}}
&&Q_2 = 
\bar s^i \ga_\mu (1 - \ga_5) d^i\ \bar s^j  \ga_\mu (1 + \ga_5)d^j 
\;, \nonumber \\
{\phantom{{l}}}\raisebox{-.16cm}{\phantom{{j}}}
&&Q_3 =  
\bar s^i  (1 + \ga_5) d^i\ \bar s^j  (1 - \ga_5)d^j 
\;,  \\
{\phantom{{l}}}\raisebox{-.16cm}{\phantom{{j}}}
&&Q_4 =  
\bar s^i  (1 - \ga_5) d^i\ \bar s^j  (1 - \ga_5)d^j 
\;, \nonumber \\
{\phantom{{l}}}\raisebox{-.16cm}{\phantom{{j}}}
&&Q_5 =   
\bar s^i \sigma_{\mu \nu} d^i\ \bar s^j   
\sigma_{\mu \nu}  d^j
\;, \nonumber 
\eea
where we use the definition in which $\sigma_{\mu\nu}=[\gamma_\mu,\gamma_\nu]/2$. 
Although the operators in the above bases
are written with both parity even and parity odd parts, only the 
parity even ones survive in the kaon matrix elements. The latter are usually parameterized 
in terms of bag-parameters $B_{{1}{\mathrm -}{5}}$, namely, 
\bea
 \label{params}
\langle \bar K^0 \vert  O_1(\nu) \vert  K^0 \rangle  &=& {8\over 3} \, m_K^2  f_K^2
\ B_1(\nu) \ ,  \nonumber \\
\langle  \bar K^0 \vert  O_2(\nu) \vert  K^0 \rangle  &=& -{5\over 3} \, \left( {m_K\over
m_s(\nu) + m_d(\nu) } \right)^2
m_K^2  f_K^2
\ B_2(\nu) \, ,  \nonumber  \\
\langle  \bar K^0 \vert  O_3(\nu) \vert  K^0 \rangle &=& {1\over 3} \, \left( {m_K\over
m_s(\nu) + m_d(\nu) } \right)^2
m_K^2  f_K^2
\ B_3(\nu)\, ,  \\
\langle  \bar K^0 \vert  O_4 (\nu) \vert K^0 \rangle &=&  2 \, \left( {m_K\over
m_s(\nu) + m_d(\nu) } \right)^2
m_K^2  f_K^2
\ B_4(\nu)\,  ,  \nonumber \\
\langle  \bar K^0 \vert  O_5(\nu) \vert   K^0 \rangle  &=& {2\over 3} \, \left( {m_K\over
m_s(\nu) + m_d(\nu) } \right)^2
m_K^2  f_K^2
\ B_5(\nu)\,  ,  \nonumber 
 \eea
where $B_i > 0$. $\nu$ in the above equation indicates the renormalization scale of the logarithmically divergent operators, $O_i$,
 and the scale at which the separation between the long-distance (matrix elements) and short-distance 
 (Wilson coefficients) physics is made. 
To make contact between the matrix elements of the operators in~(\ref{baseS}) and those in~(\ref{baseF}),  
one applies the Fierz identity on the Dirac structures (FD), which leaves 
the physical amplitude invariant, and then reorder 
the color reversed indices in operators $O_3$ and $O_5$. For example,
\bea\label{eq:f0}
&&\hspace*{-25mm}\langle \bar K^0 \vert  O_3 \vert   K^0 \rangle \equiv \langle O_3\rangle=\langle O_3^{FD}\rangle = 
\langle \bar s^i_\alpha d^j_\beta \bar s^j_\gamma d^i_\delta 
\ \left[\1\otimes\1 + \gamma_5\otimes\gamma_5\right]^{FD}_{\alpha\beta,\gamma\delta} 
\rangle = \nn \\
&&
-\frac{1}{2} \langle 
 \bar s^i_\alpha d^i_\delta \bar s^j_\gamma d^j_\beta\ \left[\1\otimes\1 + \gamma_5\otimes\gamma_5
-\sigma_{\mu\nu}\otimes \sigma_{\mu\nu}\right]_{\alpha\delta,\gamma\beta} 
\rangle = \frac{1}{2}\left( \langle Q_5\rangle - \langle Q_4\rangle\right)\,.
\eea 
Similarly, $\langle O_5\rangle=-(1/2)\langle Q_2\rangle$. Summarizing,
\bea\label{eq:f1}
&&\langle Q_1 \rangle  = \langle O_1 \rangle \;,\hspace*{11mm} \langle Q_2 \rangle  = -2 \langle O_5 \rangle \;,\cr
&&\cr
&& \langle Q_3 \rangle  = \langle O_4 \rangle \;,\hspace*{11mm} \langle Q_4 \rangle  = \langle O_2 \rangle \;, \cr
&&\cr
&&\langle Q_5 \rangle  =\langle O_2 \rangle +2\langle  O_3 \rangle \;.
\eea
In eqs.~(\ref{eq:f0},\ref{eq:f1}) and in the rest of this paper the $\nu$-dependence is 
implicit. As a side remark, we note that these formulae are strictly true only in 
the renormalization schemes in which the Dirac Fierz identity is not violated, 
such as the (Landau) $\ri$ scheme~\cite{roma-munich}.~\footnote{See ref.~\cite{sharpe} 
for a formulation of the $\msbar$ scheme in which FD is preserved at the next-to-leading 
order (NLO) in perturbation theory.} In the following we will suppose that the subtraction 
of ultraviolet divergences is made in such a renormalization scheme and restrict our 
attention to the low energy behavior of the above matrix elements.

\section{Chiral logarithmic corrections\label{sec:2}}

As mentioned in introduction, we will use ChPT to discuss the low energy behavior of 
$\Delta S=2$ operators relevant to the SUSY $\kkbar$ mixing amplitude. 
Even before considering the chiral representation of eq.~(\ref{baseS}) 
or eq.~(\ref{baseF}), it is clear that the chiral behavior of the matrix elements of  
$O_2$ and $O_3$ will be the same since these two operators differ only in the color indices. 
In other words these two operators differ by a gluon exchange, which is a local effect, that 
cannot influence the long distance behavior described by ChPT. Therefore the chiral logarithms 
in $B_2$ and $B_3$ will be the same although their respective low energy constants (LEC's) are different. 
The same argument applies to $B_4$ and $B_5$. 
This ``color blindness" is evident when working out the chiral representation of 
the operators~(\ref{baseF}). To that end, we will use the lagrangian and notation 
specified in our previous paper~\cite{ours}, and account for the following properties: 
\begin{itemize}
\item[--] Under ${\rm SU(3)_L}\otimes {\rm SU(3)_R}$ the field $\Sigma$ transforms as $\Sigma \to R\Sigma
L^\dagger$;
\item[--] The lowest order Lorentz scalars, transforming as $(\bar 3,3)$ and $(3,\bar 3)$, are 
$\Sigma$ and $\Sigma^\dagger$, respectively;
\item[--]  The lowest order Lorentz vectors, transforming as  $(8,1)$ and $(1,8)$, are 
$i\Sigma^\dagger\partial_\mu \Sigma$ and $i\Sigma\partial_\mu \Sigma^\dagger$, respectively.
\end{itemize}
We are now able to write the bosonised versions of eq.~(\ref{baseF}), namely,
\bea \label{baseC}
{\phantom{{l}}}\raisebox{-.16cm}{\phantom{{j}}}
&&Q_1 = - b_1 \frac{f^4}{8} \left( \Sigma^\dagger \partial^\mu \Sigma \right)_{ds} \left( \Sigma^\dagger \partial_\mu \Sigma \right)_{ds}
\;, \nonumber \\
{\phantom{{l}}}\raisebox{-.16cm}{\phantom{{j}}}
&&Q_2 = 
- b_2 \frac{f^4}{4} B_0^2 \Sigma^\dagger_{ds} \Sigma_{ds}
\;, \nonumber \\
{\phantom{{l}}}\raisebox{-.16cm}{\phantom{{j}}}
&&Q_3 =  
 b_3 \frac{f^4}{8} B_0^2 \Sigma^\dagger_{ds} \Sigma_{ds}
\;,  \\
{\phantom{{l}}}\raisebox{-.16cm}{\phantom{{j}}}
&&Q_4 =  
 b_4 \frac{f^4}{8} B_0^2 \left( \Sigma_{ds} \Sigma_{ds} + \Sigma^\dagger_{ds} \Sigma^\dagger_{ds} \right)
\;, \nonumber \\
{\phantom{{l}}}\raisebox{-.16cm}{\phantom{{j}}}
&&Q_5 = 
- b_5 \frac{f^4}{8} B_0^2 \left( \Sigma_{ds} \Sigma_{ds} + \Sigma^\dagger_{ds} \Sigma^\dagger_{ds} \right)
\;, \nonumber 
\eea
where we introduced the new set of bag parameters, $b_i$, with the signs chosen as to make 
all $b_i$'s positive. After sandwiching the above operators by $\langle \bar K^0\vert$ and 
$\vert K^0\rangle$ and after evaluating the matrix elements at leading order, we can relate 
$b_i$'s to the chiral limit of the $B_i$-parameters:
\bea
B_1^{\rm tree}=\frac{3}{8}b_1\,,\quad 
B_2^{\rm tree}=\frac{6}{5}b_4\,,\quad 
B_3^{\rm tree}=3\left(b_4+b_5\right)\,,\quad 
B_4^{\rm tree}=\frac{1}{2}b_3\,,\quad 
B_5^{\rm tree}=\frac{3}{2}b_2\,.
\eea  
We now proceed, like in ref.~\cite{ours}, by following the standard routine to compute the 
chiral logarithmic corrections to $B_i$. In computation of the tadpole chiral 
loop integrals we use the na\" \i ve dimensional 
regularization and the $\msbar$ renormalization scheme. Our results are:
\bea\label{chilogs}
B_1 &=& B_1^{\rm tree} \left[ 1  - \frac{1}{(4\pi f)^2}\,
\left( \frac{ m_K^2 + m_\pi^2}{2\,m_K^2} m_\pi^2\log \frac{m_\pi^2}{\mu^2} 
+ 2 m_K^2 \log \frac{m_K^2}{\mu^2}\right.\right.\nn\\
&& \left.\left. \hspace*{73.5mm}+
\frac{ 7 m_K^2 - m_\pi^2}{2\,m_K^2} m_\eta^2 \log\frac{m_\eta^2}{\mu^2} 
\right) +\dots
\right],\nn\\
B_2&=& B_2^{\rm tree} \left[ 1  - \frac{1}{(4\pi f)^2}\,
\left(  
\frac{1}{2} m_\pi^2\log\frac{m_\pi^2}{\mu^2} 
+ 4 m_K^2 \log \frac{m_K^2}{\mu^2} +
\frac{1}{6} m_\eta^2 \log \frac{m_\eta^2}{\mu^2} 
\right)  +\dots 
\right],\nn\\
B_3&=& B_3^{\rm tree} \left[ 1  - \frac{1}{(4\pi f)^2}\,
\left(  
\frac{1}{2} m_\pi^2\log\frac{m_\pi^2}{\mu^2} 
+ 4 m_K^2 \log\frac{m_K^2}{\mu^2} +
\frac{1}{6} m_\eta^2 \log\frac{m_\eta^2}{\mu^2} 
\right)  +\dots
\right],
\eea
\bea
B_4&=& B_4^{\rm tree} \left[ 1  + \frac{1}{(4\pi f)^2}\,
\left(  
\frac{1}{2} m_\pi^2\log\frac{m_\pi^2}{\mu^2} 
- 4 m_K^2 \log\frac{m_K^2}{\mu^2} +
\frac{1}{6} m_\eta^2 \log\frac{m_\eta^2}{\mu^2} 
\right)  +\dots
\right],\nn\\
B_5&=& B_5^{\rm tree} \left[ 1  + \frac{1}{(4\pi f)^2}\,
\left(  
\frac{1}{2} m_\pi^2\log\frac{m_\pi^2}{\mu^2}  
- 4 m_K^2 \log\frac{m_K^2}{\mu^2} +
\frac{1}{6} m_\eta^2 \log\frac{m_\eta^2}{\mu^2} 
\right)  +\dots
\right],\nn
\eea
where dots stand for analytic and higher order terms in ChPT, and $\mu$ is the renormalization scale.

\section{Log-safe combinations\label{sec:3}}

The phenomenological applications of the predictions based on ChPT at NLO are usually plagued by the poor 
knowledge of the size of low energy constants~\cite{CHPT}. A better predictability is then expected for 
the combinations of physical quantities in which the low energy constants cancel (partially or completely). 
Contrary to that situation, when computing the physical quantities from the 
QCD simulations on the lattice one works with light quark masses 
larger than the physical up- or down-quark ($m_{u/d}$), thus allowing one to probe 
the analytic dependence on the quark masses, while missing (again, partially or completely) the chiral 
logarithmic behavior that is expected to take over as the light quark becomes closer to physical 
$m_{u/d}$. Since the point at which the chiral logarithms, with 
coefficients predicted by $1$-loop ChPT, are to be included in extrapolations of the lattice results is 
not known, their inclusion in the chiral extrapolations induce large systematic uncertainties. 
To avoid such uncertainties one should aim at combining the physical quantities in which the 
chiral log corrections cancel. As we will show, such log-safe combinations also help reducing the 
finite volume artifacts that are becoming ever more important as the light quark gets closer to the 
chiral limit.

We now construct the {\it golden} ({\it silver}) combinations in which the chiral logarithms 
completely (partially) cancel. The criterion for creating the silver combinations will be the 
cancellation of the pion loop sum or integrals because they make the strongest deviation in 
the chiral extrapolations of the results obtained directly from lattice QCD, and because they generally
represent the most important source of the finite volume artifacts. 
From the discussion in Sec.~\ref{sec:2} and eq.~(\ref{chilogs}), one immediately identifies the 
following two golden ratios: 
\bea\label{gold1}
R_{1}^g={B_2\over B_3}\,,\quad R_{2}^g={B_4\over B_5}\,.
\eea
As for the silver combinations the simplest ones that can be deduced after inspecting eq.~(\ref{chilogs}) are  
\bea\label{silverIJ}
R_{ij}^s= B_{i} \times B_{j} \quad (i=2,3;\, j=4,5) \,.
\eea
Alternatively one can use some auxiliary quantities, preferably those that are easily calculable 
on the lattice, and combine them with bag parameters~(\ref{chilogs}) to cancel the pionic logarithms. 
Useful quantities are the decay constants and their combinations~\cite{gasser}:
\bea
f_\pi &=& f\left[ 1 - {1\over (4\pi f)^2}\left(2 m_\pi^2\log \frac{m_\pi^2}{\mu^2} + m_K^2\log
\frac{m_K^2}{\mu^2}\right)+\dots \right]\,, \nn\\
f_K &=& f\left[ 1 - {3\over 4(4\pi f)^2}\left( m_\pi^2\log \frac{m_\pi^2}{\mu^2} + 2 m_K^2\log
\frac{m_K^2}{\mu^2} + m_\eta^2\log \frac{m_\eta^2}{\mu^2}\right)+\dots \right]\,, \\
{f_K\over f_\pi} &=&  1 + {1\over 4 (4\pi f)^2} \left( 
5 m_\pi^2\log \frac{m_\pi^2}{\mu^2} 
- 2 m_K^2\log\frac{m_K^2}{\mu^2} - 
3 m_\eta^2\log \frac{m_\eta^2}{\mu^2}\right)+\dots \,, \nn \\
{f_K^2 \over f_\pi} &=& f\left[ 1 + {1\over (4\pi f)^2} \left( 
\frac12 m_\pi^2\log \frac{m_\pi^2}{\mu^2} 
- 2 m_K^2\log\frac{m_K^2}{\mu^2} - 
\frac32 m_\eta^2\log \frac{m_\eta^2}{\mu^2}\right)+\dots \right]\,. \nn
\eea
The silver log-safe combinations in which the decay constants are combined with $B$-parameters are:
\bea
R_1^s={f_K^2\over f_\pi} B_1 \,,\quad R_{2,3}^s={f_K^2\over f_\pi} B_{2,3}\,,\quad R_{4,5}^s={f_\pi\over f_K^2} B_{4,5} \,.
\eea
If, as widely expected, the $B_1\equiv B_K$ parameter is accurately determined first, other silver 
log-safe quantities are also
\bea
R_{6}^s = {B_{2,3}\over B_1}\,,\qquad R_{7}^s = {B_{4,5}\times B_1}\,.
\eea
Summarizing, in the golden ratios, $R^g_{1,2}$, all chiral logarithms cancel, 
whereas in the silver combinations, $R^s_{ij}$ and $R^s_{1-7}$, the cancellation 
of the most problematic part (from the lattice practitioner's point of view) is achieved.

\subsection{Finite volume effects}

\begin{figure}
\begin{center}
\epsfig{file=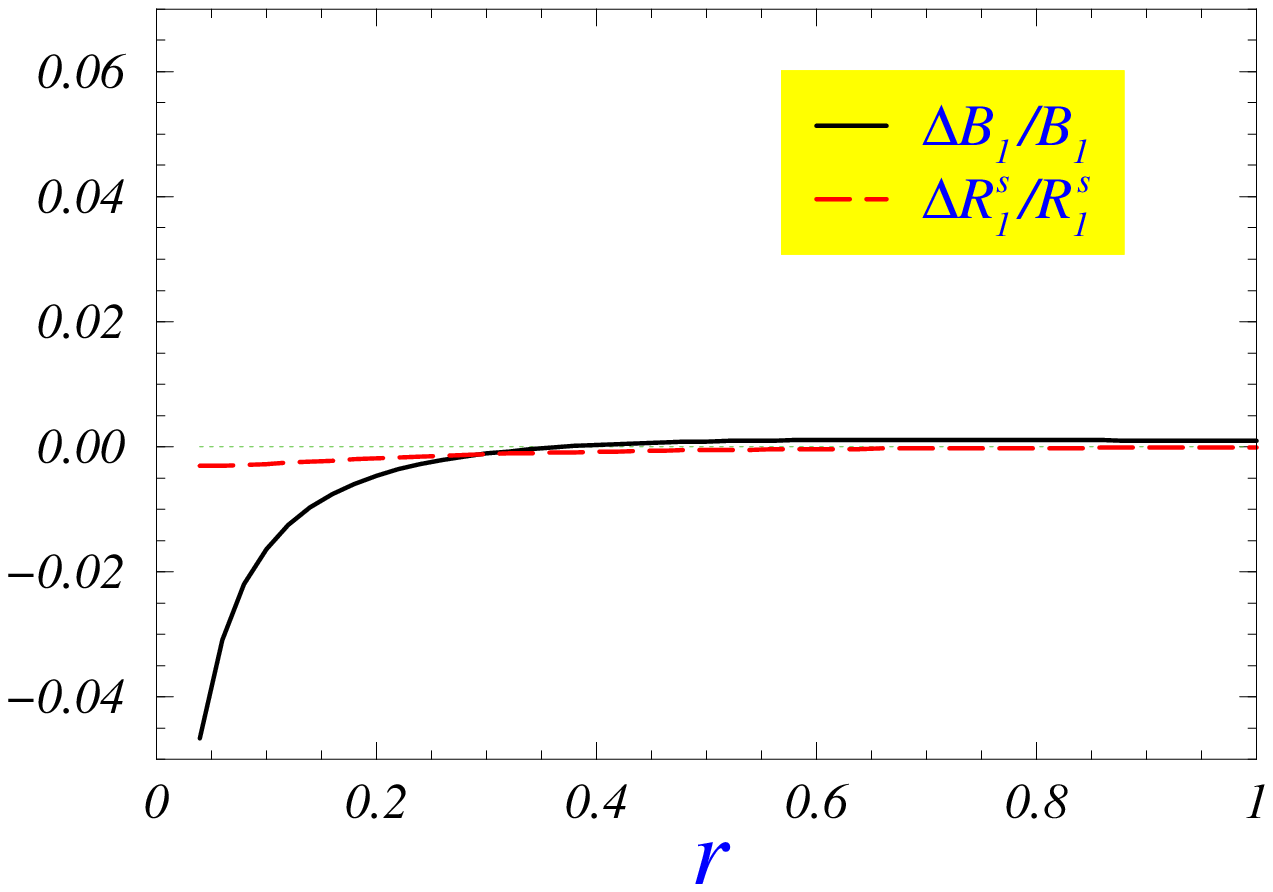, height=6.87cm} \\ 
\epsfig{file=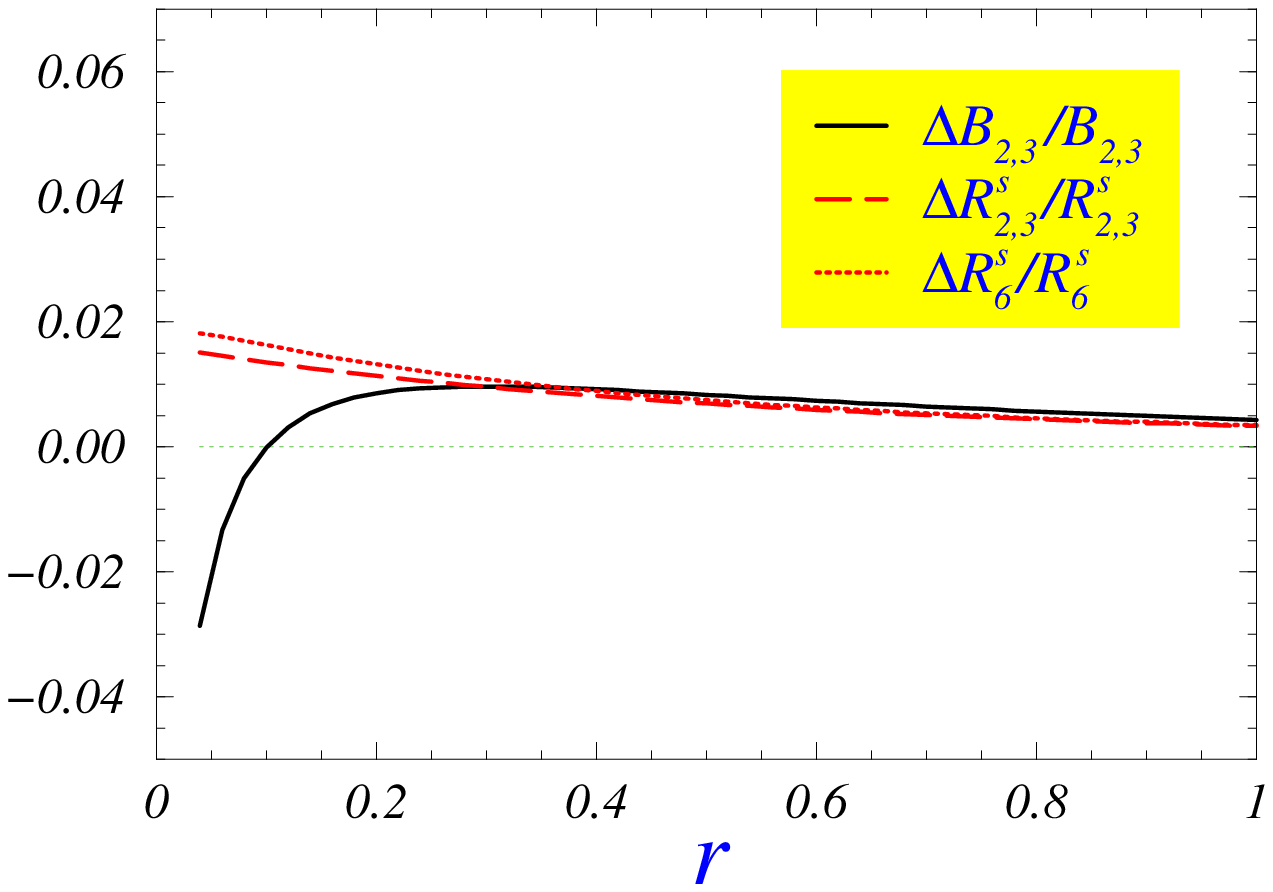, height=6.87cm} \\
\epsfig{file=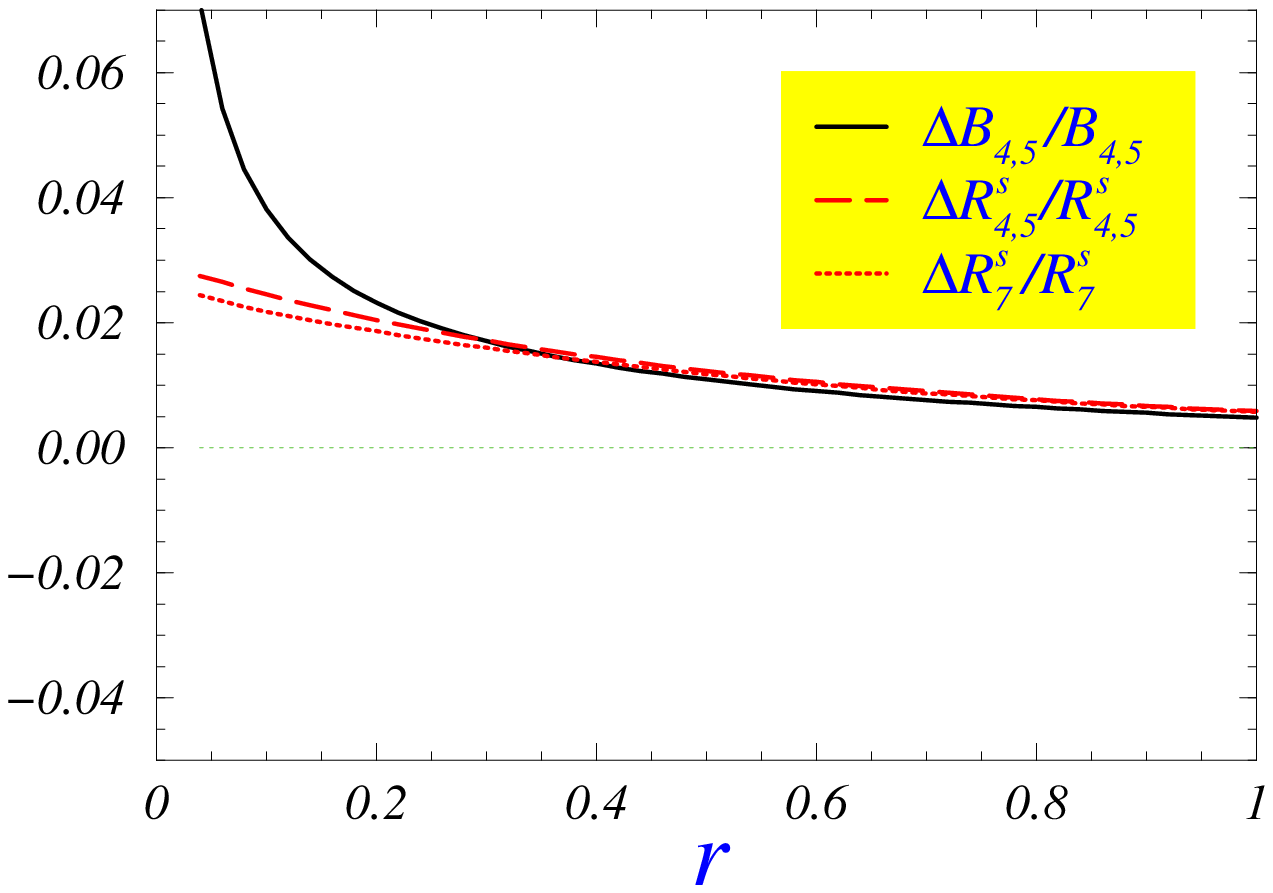, height=6.87cm}
\caption{\label{fig1}\footnotesize{\sl Finite volume effects for the $B_{1-5}$-parameters, computed by using 
eq.~(\ref{FV}), are denoted by the thick solid 
curves, whereas those corresponding to the silver combinations $R_{1-7}^s$ are 
represented by the dashed and dotted thick lines. The plots are made for $L=2\,{\rm fm}$. 
Notice that $r=m_{u/d}/m_s$ is varied by keeping $m_s$ fixed to its physical value. 
To better appreciate the benefit of using $R_{1-7}^s$ we also show 
the line corresponding to zero finite volume effect (thin dashed line).}} 
\end{center}
\end{figure}

In the finite box of side $L$, instead of integrals  
one deals with the sums over discretised momenta $\vec q=(2\pi/L)\times \vec n$ [$\vec n\in \Z^3$]. 
The difference between sums and integrals is the infrared ($\mu$ independent) effect that  
can be expressed in terms of the function ``$\xi_s$" whose properties we discussed in ref.~\cite{ours}. 
In other words, for large physical volumes, one can deduce the finite volume effects by comparing 
the expressions for a given physical quantity derived in ChPT in finite and infinite volumes~\cite{GL}.
Like in ref.~\cite{ours}, we define
\beq\label{ratioB}
{\Delta B_i\over B_i}\equiv{B_i^L - B_i^\infty \over B_i^\infty}\,,
\eeq
and obtain:
\bea\label{FV}
 {\Delta B_1\over B_1} &=& -{1\over 2 f^2}\left[ {m_K^2+m_\pi^2\over 2 m_K^2} \xi_{\frac{1}{2}}(L,m_\pi) - 
m_K^2 \xi_{\frac{3}{2}}(L,m_K) + {7 m_K^2-m_\pi^2\over 2 m_K^2} \xi_{\frac{1}{2}}(L,m_\eta)
\right],\nn\\
{\phantom{{l}}}\raisebox{-.16cm}{\phantom{{j}}}
 {\Delta B_{2,3}\over B_{2,3}} &=&  -{1\over 2 f^2}\left[ {1\over 2 } \xi_{\frac{1}{2}}(L,m_\pi) - 
2 m_K^2 \xi_{\frac{3}{2}}(L,m_K) + {1\over 6} \xi_{\frac{1}{2}}(L,m_\eta)
\right],\\
{\phantom{{l}}}\raisebox{-.16cm}{\phantom{{j}}}
 {\Delta B_{4,5}\over B_{4,5}} &= & {1\over 2 f^2}\left[ {1\over 2 } \xi_{\frac{1}{2}}(L,m_\pi) + 
2 m_K^2 \xi_{\frac{3}{2}}(L,m_K) + {1\over 6} \xi_{\frac{1}{2}}(L,m_\eta)
\right].\nn
\eea
It is important to notice that the coefficients multiplying the function $\xi_{\frac{1}{2},\frac32}(L,m_P)$ are the same 
as those in eq.~(\ref{chilogs}) multiplying $m_P^2\log(m_P^2/\mu^2)$.~\footnote{The factor 2 of mismatch between the 
coefficients multiplying the kaon part in~(\ref{chilogs}) and in~(\ref{FV}) is canceled by a factor of two in the 
definition of $\xi_{\frac{3}{2}}(L,m_K)$. } 
Therefore the combinations of physical quantities in which the chiral logarithms cancel 
 not only allow for the safer chiral extrapolations of their lattice estimates,  but 
they also provide the cancellation of the finite volume effects (at least those 
predicted by the $1$-loop ChPT). For the silver combinations there is however a subtlety: 
although exponentially suppressed [$\propto~\hspace{-8pt}\exp(-m_K L)$], the terms proportional 
to $\xi_{\frac32}(L,m_K)$ are numerically important because they have larger factors 
in front and because the function $\xi_{\frac32}$  has worse 
infrared behavior than $\xi_{\frac12}$. Therefore it is also important to be careful in dealing 
with terms corresponding to the kaon loops. Of all the silver combinations discussed in this section, 
only $R^s_{ij}$ [see eq.~(\ref{silverIJ})] receive large finite volume corrections while all the other 
silver combinations do not suffer from this problem. The finite volume effects for $R^s_{1-7}$, together 
with those for $B_{1-5}$, are plotted in fig.~\ref{fig1}.
We first observe the usual behavior, namely that the finite volume effects become larger 
as one is getting closer to the physical point, i.e., $r=0.04$~\cite{leutwyler}. 
Secondly, we see that the finite volume effects on the silver combinations, 
$R_{1-7}^s$, are clearly reduced when confronted to those that plague $B_{1-5}$-parameters. 
Finally, we stress again that the leading finite volume effects on the golden ratios 
[eq.~(\ref{gold1})] are totally absent.

\section{$K\to \pi$ case \label{sec:4} }

The discussion of the previous section can be easily extended to the $K\to \pi$ case which is often 
considered on the lattice when computing the $\Delta I=3/2$ amplitude in $K\to \pi\pi$ decay. 
A golden log-free combination can be easily constructed from the chiral logarithmic corrections to the 
electro-weak penguin operators calculated in, for example, ref.~\cite{cirigliano}. 

To be more specific we will concentrate on the following three operators: 
\bea
Q_7^{3/2}&=&\bar s\gamma_\mu^L u \,\bar u\gamma_\mu^R d + \bar s\gamma_\mu^L d \, (\bar u \gamma_\mu^R u -\bar
d \gamma_\mu^R d ), \nn\\
Q_8^{3/2}&=&\bar s^i\gamma_\mu^L u^j \, \bar u^j\gamma_\mu^R d^i + \bar s^i\gamma_\mu^L d^j \, (\bar u^j
\gamma_\mu^R u^i -\bar d^j \gamma_\mu^R d^i ), \\
Q_{27}^{3/2}\equiv Q_{9}^{3/2}&=&\bar s\gamma_\mu^L u \, \bar u\gamma_\mu^L d + \bar s\gamma_\mu^L d \, (\bar u \gamma_\mu^L u -\bar d
\gamma_\mu^L d ) ,\nn
\eea
where $\gamma_\mu^{R/L}=\gamma_\mu(1\pm\gamma_5)$. The chiral representation of these operators reads
\bea
Q_7^{3/2}&=&-b_2 {f^4\over 4}B_0^2 \left( \Sigma^\dagger_{du}\Sigma_{us} 
- \Sigma^\dagger_{dd}\Sigma_{ds} + \Sigma^\dagger_{uu}\Sigma_{ds}\right),\nn\\
Q_8^{3/2}&=&-b_3 {f^4\over 4}B_0^2 \left( \Sigma^\dagger_{du}\Sigma_{us} 
- \Sigma^\dagger_{dd}\Sigma_{ds} + \Sigma^\dagger_{uu}\Sigma_{ds}\right),\\
Q_{27}^{3/2}&=&-b_1 {f^4\over 8} \left\{ (\Sigma^\dagger\partial_\mu\Sigma)_{ds} 
[  (\Sigma^\dagger\partial_\mu\Sigma)_{uu}-(\Sigma^\dagger\partial_\mu\Sigma)_{dd}] +
(\Sigma^\dagger\partial_\mu\Sigma)_{us} (\Sigma^\dagger\partial_\mu\Sigma)_{du} \right\}.\nn
\eea
The relevant matrix elements are parameterized as
\bea
\langle \pi^+\vert Q_7^{3/2}\vert K^+ \rangle &=&-{2\over 3} \, \left( {m_K\over
m_s  + m_d  } \right)^2
m_K^2  f_K^2
B_5^{K\pi},\nn\\
\langle \pi^+\vert Q_8^{3/2}\vert K^+ \rangle &=& - 2 \, \left( {m_K\over
m_s  + m_d  } \right)^2
m_K^2  f_K^2
B_4^{K\pi},\\
\langle \pi^+\vert Q_{27}^{3/2}\vert K^+ \rangle &=&{4\over 3} \ 
m_K m_\pi  f_K^2
B_1^{K\pi},\nn
\eea
where the bag parameters $B_{1,4,5}^{K\pi}$ are the $K\to\pi$ versions of the $\kkbar$ ones,  
discussed in the previous section. They both have the same tree level values but differ 
in the chiral logarithmic corrections, which we computed as well and obtained
\bea\label{chilogsKPI}
B_1^{K\pi} &=& B_1^{\rm tree} \left[ 1  - \frac{1}{4 (4\pi f)^2}\,
\left( \frac{ 5m_K -13 m_\pi}{m_K-m_\pi} m_\pi^2\log \frac{m_\pi^2}{\mu^2} 
+ 2 \frac{ 3 m_K + m_\pi}{m_K-m_\pi} m_K^2 \log \frac{m_K^2}{\mu^2}\right.\right.\nn\\
&& \left.\left. \hspace*{78.5mm}-
3 m_\eta^2 \log\frac{m_\eta^2}{\mu^2} 
\right) +\dots
\right],\nn\\
B_4^{K\pi} &=& B_4^{\rm tree} \left[ 1  + \frac{1}{4 (4\pi f)^2}\,
\left( \frac{ 3m_K +5 m_\pi}{m_K-m_\pi } m_\pi^2\log \frac{m_\pi^2}{\mu^2} 
+ 2 \frac{  m_K -5 m_\pi}{ m_K-m_\pi } m_K^2 \log \frac{m_K^2}{\mu^2}\right.\right.\nn\\
&& \left.\left. \hspace*{78.5mm}-
\frac{7}{3} m_\eta^2 \log\frac{m_\eta^2}{\mu^2} 
\right) +\dots
\right],\\
B_5^{K\pi} &=& B_5^{\rm tree} \left[ 1  + \frac{1}{4 (4\pi f)^2}\,
\left( \frac{ 3m_K +5 m_\pi}{m_K-m_\pi } m_\pi^2\log \frac{m_\pi^2}{\mu^2} 
+ 2 \frac{  m_K -5 m_\pi}{ m_K-m_\pi } m_K^2 \log \frac{m_K^2}{\mu^2}\right.\right.\nn\\
&& \left.\left. \hspace*{78.5mm}-
\frac{7}{3} m_\eta^2 \log\frac{m_\eta^2}{\mu^2} 
\right) +\dots
\right] .\nn
\eea
Therefore, the chiral extrapolation of the ratio of matrix elements of the electro-weak penguin operators, i.e., 
\bea
R_2^{g \ K\pi}={B_4^{K\pi}\over  B_5^{K\pi}}\,, 
\eea
that can be computed on the lattice, is free from uncertainties induced by the inclusion of the chiral logarithms in chiral
extrapolations. In addition, the leading finite volume effects in the ratio $R_2^{g \ K\pi}$ completely cancel.   
The absolute values for $B_{1,4,5}^{K\pi}$, on the other hand, can be extracted from the following 
silver relations:
\beq
R^{s \ K\pi}_1=\frac{f_K}{f_\pi}B_1^{K\pi}\,, \qquad R^{s \ K\pi}_{4,5}={f_K}\,B_{4,5}^{K\pi} \,.
\eeq
We believe the safe extrapolation of the lattice results for $R_2^{g \ K\pi}$  may be useful in 
disentangling the current discrepancies among various analytic~\cite{knecht} and lattice 
approaches~\cite{lattice} used so far to estimate $\langle \pi^+\vert Q_8^{3/2}\vert K^+ \rangle$. 
Even more so after noticing that almost all approaches agree in the value for $B_5^{K\pi}$, 
while they differ quite a lot in that for $B_4^{K\pi}$. Finally, a reader interested in the discussion 
of the potential SUSY enhancement of the $\Delta I=3/2$ amplitude is referred to ref.~\cite{neubert}.

\section{Golden ratios and next-to--next-to leading (NNLO) chiral corrections\label{sec:6}}

We showed that the golden ratios were completely protected against the presence of the chiral logarithms and thus also 
against the finite volume corrections. One may wonder if that feature survives after accounting for higher order corrections 
in the chiral expansion. That question was recently rised in ref.~\cite{durr} where it has been argued that 
the pion mass might receive sizable finite volume NNLO chiral corrections. That issue will obviously 
be settled only after the full $2$-loop chiral corrections in finite volume are computed. Here we discuss 
such corrections to our golden ratios $R_{1,2}^g$, and argue that in this case the finite volume corrections 
arising beyond $1$-loop ChPT are indeed negligible.
Golden ratios are generically defined as 
\bea
R^g={\langle out\vert {\cal O}_A\vert in\rangle\over \langle out\vert {\cal O}_B\vert in\rangle}\,,
\eea
where ${\cal O}_A$ and ${\cal O}_B$ are two different operators with the same chiral representation. The NNLO chiral 
expansion to this ratio can be schematically written as 
\bea
R^g&=&\frac{g_A [1+(logs+c_A)+(dlogs+clogs_A+C_A)+\dots ]}{g_B [1+(logs+c_B)+(dlogs+clogs_B+C_B)+\dots ]}  \\
&=&\frac{g_A}{g_B}\biggl\{ 1+(c_A-c_B)+\left[(clogs_A-clogs_B)+(c_B-c_A)(c_B+logs)+(C_A-C_B)\right]+\dots \biggr\},\nn
\eea
where $g_A/g_B$ is the tree level value of the golden ratio, $c_{A,B}$ and $C_{A,B}$ are the low energy constants arising at 
 NLO and NNLO respectively, $clogs_{A,B}$ are the $1$-loop part to the NNLO
 correction. Based on the same argument used 
in sec.~\ref{sec:2}, the single and double chiral logarithms, $logs$ and $dlogs$, completely cancel in the ratio and thus  
the computation of the NNLO chiral corrections comprises the $1$-loop diagrams only. 
That feature clearly holds at any other higher order, i.e. one needs to compute the diagrams that are one 
loop less with respect to the ones indicated by the canonical ChPT counting.

After including the NNLO corrections, the finite volume effects to the golden ratio can be written in the following form:
\bea
{\Delta R^g\over R^g}\equiv {R^{gL}-R^{g\infty}\over R^{g\infty}} =
\Delta clogs_A - \Delta clogs_B + (c_B-c_A) \Delta logs\,,
\eea 
in an obvious notation. We see that the finite volume effects arise only from
the $1$-loop diagrams at the weak vertices, and therefore the result can be 
expressed in terms of the $\xi_s$-function, as before.

To exemplify the above discussion we now use the set of counterterm lagrangians, written explicitely in 
ref.~\cite{cirigo}, to obtain the counterterm contributions to the NLO expression for $R_2^g$. Those are  
then used to calculate the finite volume corrections to $R_2^g$ at NNLO. We obtain~\footnote{To easily identify the counterterm
coefficients from ref.~\cite{cirigo}, our $\delta$'s and their $c$'s are related as
\bea
\delta_i&\equiv&\frac{c_i^{(4)}}{g_4}-\frac{c_i^{(5)}}{g_5},\nonumber
\eea
with $g_4=2B_0^2 f^2 B_4^{\rm tree}$, and $g_5=(2/3)B_0^2 f^2 B_5^{\rm tree}$.}
\bea
{\Delta R^g_2\over R^g_2}&=&-
\frac{1}{12\,f^2} \l \{ 12\,m_K^2\,\left[ {{\delta }_1} + 3\,{{\delta }_2} + 
       2\,\left( -3\,{{\delta }_3} + 8\,{{\delta }_4} + 8\,{{\delta }_5} + 8\,{{\delta }_6} - 4\,{{\delta }_7} \right)  \right] \,
     \xi_\frac12 (L,{m_K})\r. \nn \\
	 &&+ 3\,\left[ 8\,m_K^2\,{{\delta }_2} - 
       {{m_{\pi }}}^2\,\left( {{\delta }_1} + {{\delta }_2} + 
          4\,\left( 3\,{{\delta }_3} - 5\,{{\delta }_4} - 5\,{{\delta }_5} - 6\,{{\delta }_6} + 6\,{{\delta }_7} \right)  \right) 
       \right] \, \xi_\frac12 (L,{m_{\pi }}) \nn \\ 
	   && + \,\left[ 
        32\,m_K^2\,\left( {{\delta }_4} + {{\delta }_5} + {{\delta }_6} \right) 
	- 4  
          m_\pi^2\,\left( 3\,{{\delta }_4} + 3\,{{\delta }_5} + 2\,{{\delta }_6} \right)   \r. \nn \\
		&&\qquad \l.\l. - 
		m_\eta^2\,\left( {{\delta }_1} + 17\,{{\delta }_2} + 20\,{{\delta }_3} + 24\,{{\delta }_7} \right)  \right] \,
     \xi_\frac12 (L,{m_{\eta }}) \r\}.
     \eea
Notice that the double pole contributions ($\propto \xi_{\frac{3}{2}}$) coming from NLO$^2$
and from NNLO terms, cancel against each other. The size of the finite volume corrections evidently depends 
on the values for the low energy constant $\delta_i$. By using either the na\" \i ve dimensional analysis or the values extracted 
from the recent quenched lattice data~\footnote{We thank Mauro Papinutto for sending us the lattice estimates 
for $\delta_i$'s prior to their publication.}, we get that for $L\geq 1.5$~fm and $r\in [0.04,1)$, the finite volume effects, 
$\Delta R^g_2/R^g_2$, are within a few per mil, thus totally negligible.

\section{Summary\label{sec:5}}

In this work we computed the $1$-loop chiral corrections to the bag parameters that are relevant to the 
 $\kkbar$ mixing amplitude in the SUSY extensions of the SM. Two out of five independent 
 $\Delta S=2$ operators have reversed color indices so that their 
 chiral log corrections are identical to those in which the color indices are not reversed. 
 This ``color blindness", which also emerges from the explicit calculation in the chiral 
 representation of those  $\Delta S=2$ operators, is actually expected since 
the change of the color structure of the operators is a local effect which cannot appear 
in chiral loops. Rather it is encoded in the low energy constants. 
After comparing the $1$-loop chiral expressions for all the bag parameters obtained in finite and 
infinite volume, we also provide the formulae for the finite volume effects which are 
useful for the assessment of the associated systematic uncertainty in the lattice calculations.
As in our previous paper~\cite{ours}, we see that the finite volume effects become more pronounced as 
the light valence quark of the neutral kaon gets closer to the chiral limit. We find that the combinations 
of bag parameters in which the chiral logarithmic dependence cancels partly or totally may be
particularly beneficiary for the lattice calculations because they allow one to sensibly reduce two 
important sources of systematic uncertainties: (i) the errors induced by the chiral extrapolations, 
and (ii) the errors due to the finiteness of the lattice box.
We construct the explicit combinations of bag parameters that are log-safe, i.e., in which the 
chiral logarithms cancel either totally (golden combinations, $R^g_{1,2}$) or partially (silver combinations, $R^s_{1-7}$).
In addition, we show that inclusion of the NNLO corrections does not spoil the advanteges of computing the golden ratio 
on the lattice. A similar discussion is also made for the case of the $K\to \pi$ matrix elements of the electro-weak 
penguin operators.

\vspace*{1.cm}

\section*{Acknowledgment}

We thank G.~Isidori, V.~Lubicz and G.~Martinelli for comments on the manuscript.

\end{document}